# A Dual Framework for Optimized Data Storage and Retrieval using Lightweight Python Blockchain and Scalable Smart Contracts with IPFS

Vatsala Upadhyay[1], Dr. J. Kokila[2], Dr. Abhishek Vaish[3*]

[1]Department of Information Technology, Indian Institute of Information Technology Allahabad, Prayagraj, 211015, Uttar Pradesh, India. [2]Department of Computer Science and Engineering, Indian Institute of Information Technology Trichy, Tiruchirappalli, 620012, Tamil Nadu, India.
[3]Department of Information Technology, Indian Institute of Information Technology Allahabad, Prayagraj, 211015, Uttar Pradesh, India.

*Corresponding author(s). E-mail(s): abhishek@iiita.ac.in;
Contributing authors: vatsau01@gmail.com; kokilaj@iiitt.ac.in;

## Abstract

The exponential growth of IoT data demands efficient, secure, and scalable storage solutions on one hand, and efficient data migration and retrieval on the other hand are essential for the systems to be practical and acceptable for different applications. The traditional cloud-based models face latency, security, and high operational costs, while existing bi-directional data storage and retrieval-based IPFS models are not computationally efficient and incur high gas costs at the cost of a necessary blockchain deployment. To overcome the challenges of efficient data migration, we initially developed a 2-way data storage and retrieval system as well as a scalable framework that dynamically monitors and transfers device-generated data to IPFS, records the content identifier(CID) on a blockchain, and enables secure, real-time access via smart contracts. Experimental results demonstrate that the existing work achieved an average data upload time of 117.12 sec for a file size of 500 MB; our framework achieves a faster upload time of 7.63 sec, marking a 93.47% improvement. We further optimize the proposed framework to reduce the file upload time incurred from the smart contracts by introducing a blockchain-inspired, lightweight, and customizable Python framework that replicates the storage and retrieval functionalities of a traditional blockchain, where the file upload time is 4.2 sec, further optimized by 45% from our previous approach, thus demonstrating its efficiency, security and suitability for deploy ment in real-time and critical IoT applications and outperforming the existing IPFS-smart contract based solutions.

## 1 Introduction

The IoT has transformed various industries by enabling connectivity and automation. However, as the number of devices grows, managing and storing the volume of generated data presents challenges that demand solutions for scalable, secure, and efficient data storage. IoT devices operate with constraints on storage, computing power, network bandwidth, and energy, raising the concern of efficient data management[1]. In addition, IoT devices' heterogeneity, dynamic nature, and limited computational capabilities complicate the storage issue[2]. Ensuring real-time processing, security, and privacy for sensitive IoT data is another major challenge, as unauthorized access to stored information can compromise system integrity[3, 4]. While cloud storage is often used for IoT data management, it suffers from latency issues, high operational costs, and centralized control, limiting its effectiveness in meeting the real-time and low-latency requirements of critical IoT applications[5]. For instance, blockchain and IPFS-based service models have been proposed as alternatives, but their integration with IoT still faces challenges related to seamless data retrieval, 2-way communication, and optimizing storage efficiency[5]. Existing IoT storage solutions fall into categories like Local storage, where data are stored directly on the device and are efficient for small-scale applications but unsuitable for large data sizes due to hard ware limitations[6]; Edge storage, where the data is stored at the network edge to reduce the bandwidth and latency requirement with the trade-off in advanced pro cessing capabilities for the edge devices[7]. Cloud storage, where the cloud servers are used for data storage and analysis for offering scalability[8], and Decentralized solutions like the Blockchain and IPFS have emerged as possible alternatives for the IoT storage solutions that also ensure data immutability, transparency[9, 18]. Hybrid storage that offers a combination of centralized storage (like the cloud) and decentralized storage to seek their benefits of providing security and enhanced performance as in [10]. Several researchers have explored solutions like Centralized and Decentralized solutions for optimizing the data storage and ensuring data scalability and security[11]. Centralized storage mechanisms, like the cloud-based solutions, rely on architectures where the data is transmitted to a remote cloud server for data processing and retrieval. The model offers scalability and centralized data management at the cost of data breaches, single points of failure, the need for high operational costs, and high latency that is not feasible for real-time applications[12, 13] and scenarios for retrieval of data from devices located in remote areas[14]. Also, the centralized cloud storage solutions do not support edge-computing[15], thus prompting hybrid storage architectures that combine cloud computing and decentralized models for enhanced reliability and security[16].

On the other hand, Decentralized solutions like Blockchain and IPFS enhance fault tolerance and the elimination of single-point failures[17] in addition to offering a secure and transparent method of managing the data without relying on a centralized entity. Blockchain and IPFS also enhance data integrity and mitigate security risks[6]. While Blockchain strengthens the data integrity using the smart-contract-based control[18], IPFS has been chiefly adopted as a file-sharing protocol for efficient data retrieval and data storage in a distributed manner, making it suitable for data caching[9].

IPFS, as a decentralized solution, is a suitable solution for scalable IoT storage and its feature of content-addressable storage that reduces latency and increases data redundancy[4, 19]. IPFS allows data distribution across multiple nodes in a network for the elimination of single-point failures[20]. Additionally, integrating blockchain with IPFS provides immutability and security to the stored data, reduces storage costs, and resilience to cyber threats[19], enabling tamper-proof storage for the IoT data[4, 8]. This integration has proved useful in applications like healthcare data, vehicular networks, wireless sensor networks, and smart agriculture[2, 18, 20, 21], where real-time data access and data management are critical. While IPFS offers a scalable and distributed storage model, it primarily functions as a read-only system for IoT devices, lacking seamless real-time updates and efficient retrieval mechanisms[19, 20]. The need for a 2-way approach for data storage and retrieval in decentralized IoT storage arose from the limitations of traditional cloud-based models, which suffer from high latency, security vulnerabilities, and dependency on centralized control. To address these gaps, several studies have explored the integration of IPFS with blockchain-based smart contracts to enable a 2-way communication approach between IoT devices and storage networks, enhancing data integrity, security, and accessibility[4, 19]. In this approach, IoT devices upload data to IPFS, generating a content identifier(CID) that uniquely references the stored data. This CID is then stored on a blockchain smart contract, ensuring integrity, authenticity, and accessibility without relying on a centralized authority[17]. An integrated blockchain and IPFS model for secure data storage was introduced in [4] using a middleman approach, where the smart contracts mediate data interactions for enhancing efficiency and retrieval accuracy. Similarly, in [22], a secure file-sharing system was developed by leveraging IPFS and blockchain to provide immutable, traceable, and decentralized storage, enabling reliable 2-way data transactions between the IoT devices and the storage nodes. Beyond storage, access control and authentication mechanisms have also been integrated into the frameworks, as in [23], where a blockchain-based authentication model for secure IPFS-based file storage is created to ensure that only authorized entities can retrieve stored data. By leveraging 2-way communication between the IoT device and IPFS, the approach can offer efficient, secure, and scalable data storage and retrieval mechanisms. Using blockchain smart contracts for CID management, access control, and authentication ensures tamper-proof, decentralized, and dynamically updatable data storage[17, 24]. However, there are certain drawbacks in the existing 2-way communication approaches using IPFS related to efficiency, security, and scalability. For instance, in [17], it is seen that while IPFS enhances decentralized storage, it suffers from high retrieval latency and inefficient content indexing, which hinders real-time communication in IoT environments. Similarly, in [9], it is pointed out that keyword-based search mechanisms in IPFS-integrated vehicular networks introduce computational overhead, making real-time data retrieval less efficient. Also, smart-contract-based IPFS storage increases blockchain transaction costs, limiting the scalability of the 2-way communication[18]. Further, integrating the consensus mechanism with IPFS contributed to an increased computational burden, which can be a bottleneck for lightweight IoT devices[4]. Authentication challenges can also arise in fog computing environments[25], where integrating IPFS with blockchain introduces delays in verifying and retrieving stored data. The storage constraint in IoT due to IPFS's reliance on distributed node availability can lead to data loss or retrieval failures[26]. In [27], a 3-tier storage framework that attempts to mitigate IPFS's security vulnerabilities is presented, but privacy concerns persist due to IPFS's immutable nature, making it difficult to update or delete sensitive information.

These limitations indicate the need for further optimization of IPFS-based 2-way storage and retrieval frameworks to enhance efficiency, reduce retrieval latency, and improve security in resource-constrained environments.

1.1 Our contributions

Based on the analysis of the existing storage solutions for the IoT, we propose the optimized data storage and retrieval model in progressive phases: the first phase is an optimized and scalable 2-way data storage and retrieval framework between the IoT device and IPFS for efficient data storage, retrieval, and scalability using smart contracts that enable real-time file monitoring and dynamic retrieval via a smart contract and an IPFS gateway, and the second phase is the further optimized version of the first phase that further reduces the latency and gas costs incurred through a blockchain-inspired lightweight Python blockchain framework as an alternative to the traditional blockchain that consumes heavy computational load. Figure 1 shows the major contributions of this work, which have also been listed below:

1. A blockchain-inspired customizable Python blockchain to store the CIDs and for efficient data storage and retrieval with low memory usage and faster upload and retrieval time for secure data management.
2. A SHA-256-based hash verification mechanism for ensuring tamper-proof data storage using the lightweight Python-blockchain framework, tailored for IoT edge devices like sensors and wearable devices.
3. A bi-directional model that enables real-time transfer between the device and IPFS for automating data storage and retrieval using smart contracts for reducing manual intervention and enhancing the system response.
4. Introducing a real-time, automated mechanism to dynamically detect the created files on the device(s) and immediately upload them to IPFS, and storing the CIDs obtained from IPFS directly in an Ethereum smart contract for secure, immutable tracking of stored files.

Fig. 1: The major contributions of the proposed optimized data storage and retrieval frameworks.

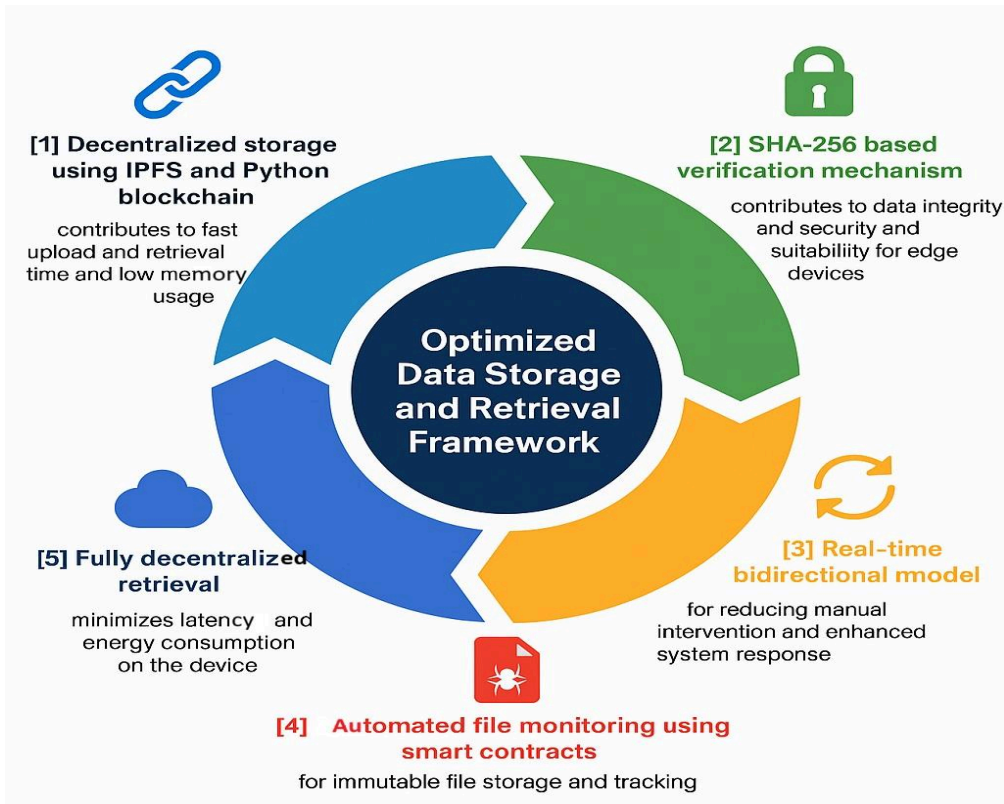

5. A fully decentralized retrieval mechanism(using IPFS and smart contracts) that minimizes the latency and dependency on centralized servers(the device) and optimizes the storage efficiency, latency, and power consumption to reduce the overhead on the device.

The rest of the paper is organized as follows: Section 2 discusses the Literature review done for the various centralized, decentralized, and other storage solutions for the IoT, the methodologies and shortcomings in the existing works; Section 3 discusses the Experimental setup, dataset, and hardware specifications; Section 4 describes the Results and Analysis obtained from the experiment performed and a comparison with the existing works. Finally, Section 5 discusses the summary of our work and the scope of further improvement in the resultant work for the research fraternity to explore.

## 2 Literature Review

Decentralized storage solutions combining blockchain and the IPFS have emerged as promising for secure and efficient data storage in the IoT environment. Several studies have proposed frameworks combining these technologies to address data privacy, scalability, integrity, and low-latency access. For instance, a low-cost storage scheme for IoT data has been proposed in [2] that uses tiered storage using blockchain and IPFS. This method reduces storage costs while maintaining real-time access. However, its effectiveness is compromised by heavy data loads and dynamic updates.

In [3], the authors explore agricultural IoT data storage optimization using blockchain technology. By integrating IPFS for decentralized storage, the system ensures data integrity and security in agricultural applications. The approach offers tamper-proof data records and enhanced traceability. However, the implementation in rural areas may face challenges such as limited network infrastructure and resource limitations of IoT devices.

Similarly, in [4], the authors introduce the integration of a lightweight consensus mechanism and IPFS for enhancing IoT data privacy. The approach is suitable for deployment in large-scale applications. The scheme reduces the computational burden through improved energy efficiency but suffers from reduced transaction throughput. Also, the system may become vulnerable to attacks as the consensus mechanism isn't robust and provides fewer security guarantees.

The authors in [6] propose an efficient decentralized storage scheme using the combination of public blockchain and IPFS. This approach improves data availability by distributing the encrypted IoT data across multiple IPFS nodes. This ensures data redundancy and fault tolerance, minimizing the risk of data loss. The disadvantage of the method lies in the resource-intensive nature of IPFS node maintenance and blockchain synchronization for a resource-constrained environment like IoT.

In [7], the authors propose an efficient big-data storage system for a distributed IoT environment. The architecture optimizes block size and IPFS chunking for improved performance. While the method reduces data duplication, it suffers from high processing overhead when handling frequent data updates that limit real-time applications.

Similarly, in [8], the authors have designed a service model combining blockchain and IPFS for IoT data sharing. The solution emphasizes access control and data encryption, thus providing robust privacy. However, the resource-intensive encryption operation hinders its feasibility on low-power IoT devices.

The authors in [9] propose a storage system for vehicular data using the blockchain and IPFS with keyword search capabilities. The keyword search allows efficient data retrieval while maintaining user privacy through encrypted metadata indexing. The advantage of the method lies in facilitating fast and secure access to large-scale vehicular data. However, scalability issues arise due to the growing size of the metadata stored on the blockchain, increasing the energy and operational cost, thus impacting long-term performance.

A scalable data-sharing method for smart agriculture using blockchain-based smart contracts has been proposed in [10] that ensures secure data exchanges and promotes trust among stakeholders. Limited throughput and latency bottlenecks restrict the application to scenarios demanding real-time monitoring, a disadvantage of this approach.

A framework for malicious node detection for the Wireless Sensor Networks(WSN) has been proposed in [12] using IPFS and blockchain. The architecture leverages reputation scoring for node verification by enhancing data authenticity. However, the system faces challenges in scaling due to the increasing complexity of managing large networks.

Similarly, in [13], the authors have introduced a hierarchical classified storage and incentive consensus scheme for building IoT systems using blockchain. The method categorizes data based on sensitivity and employs an incentive mechanism to encourage data sharing and storage. The hierarchical approach enhances data management efficiency and security. However, designing effective incentive models can be complex and require continuous adjustments to align with user behavior.

In [16], the authors have proposed a new model named BSSN, a system that enables adjustable blockchain storage tailored for resource-constrained IoT scenarios. BSSN optimizes resource usage and enhances system flexibility by allowing dynamic storage allocation. This adaptability is beneficial for IoT devices with limited storage capacities. However, the complexity of managing adjustable storage may lead to implementation challenges and potential security risks if not properly managed.

In [17], the authors present a distributed data storage and retrieval system using the IPFS and blockchain to ensure secure and tamper-proof data management. The system uses content addressing in IPFS and transactional immutability in blockchain to protect against unauthorized data modification. The approach improves data integrity and reduces storage overhead by keeping only the hashes on-chain and storing the actual data off-chain. However, the reliance on the consensus mechanism can lead to an increase in the latency and transaction cost for real-time IoT applications.

Similarly, in [18], the authors present a smart-contract-based IoT storage framework using Ethereum and IPFS for automated data verification and retrieval. Smart contracts enhance data transparency and automate access control. The method also ensures decentralized verification and secure sharing, but the overhead of executing smart contracts leads to higher gas costs and latency in low-resource IoT devices.

A healthcare data management framework in [19] that integrates blockchain and IPFS for secure patient record sharing has been proposed. This solution ensures data confidentiality and patient privacy at the cost of interoperability issues with the existing healthcare systems, thus limiting this proposed work.

In [20], the authors have proposed a trustworthy IoT data streaming using blockchain and IPFS. This approach ensures data provenance and real-time auditability, essential for mission-critical IoT applications. However, the bandwidth consumption increases due to the continuous synchronization of IoT streams.

The authors in [25] have introduced a smart-contract-based storage scheme for a fog computing environment that facilitates secure authentication and device integrity. The limitation of the work lies in the high latency incurred when verifying data access across multiple fog nodes.

A hybrid architecture combining decentralized data storage with centralized management has been proposed in [26]. This structure leverages blockchain and IPFS for data distribution while maintaining centralized oversight for administrative tasks. The method offers improved scalability and data redundancy. However, the dual structure introduces complexity in system integration and potential latency due to the coordination between centralized and decentralized components.

In [28], the authors proposed a decentralized access model for IoT data using a modular consortium architecture that integrates blockchain and IPFS. This approach enhances data privacy by eliminating centralized control and reducing single-point failures. However, the complexity of managing a consortium blockchain can lead to interoperability challenges and increased implementation costs. computational overhead on constrained devices.

**Table 1: Comparison of Blockchain and IPFS-Based IoT Data Storage Techniques.**

| Name of Technique | Dataset Used | Parameters Considered for Evaluation | Type of Blockchain Used | Drawbacks in the Approach |
|---|---|---|---|---|
| Agricultural IoT Data Storage [3] | Agricultural IoT datasets | Data Integrity, Traceability | Public Blockchain, IPFS | Network infrastructure limitations in rural areas |
| Lightweight Consensus Framework [4] | – | Scalability, Energy Efficiency | Permissioned Blockchain, IPFS | Reduced security in lightweight consensus |
| Medical IoT Storage Framework [5] | Medical sensor data | Privacy, Data Integrity | Private Blockchain, IPFS | High overhead for maintaining privacy-preserving computations |
| Efficient Decentralized Storage [6] | Public IoT datasets | Data Redundancy, Retrieval Speed | Public Blockchain, IPFS | High overhead for maintaining data consistency |
| Big Data Retrieval System [7] | Large-scale big data | Data Access Speed, Storage Cost | Public Blockchain, IPFS | Scalability limitations with increasing data volume |
| Smart Contract for Halal Certification [8] | Certification data | Immutability, Traceability | Public Blockchain, IPFS | High transaction fees, scalability issues |
| Smart Contract for Smart Agriculture [10] | Smart agriculture datasets | Data Integrity, Sharing Efficiency | Public Blockchain, IPFS | Latency and transaction costs |
| Massive Data Storage Model [11] | Large-scale IoT datasets | Data Availability, Fault Tolerance | Public Blockchain, IPFS | Synchronization delays with increasing blockchain size |
| WSN for Malicious Node Detection [12] | Wireless Sensor Network (WSN) data | Security, Detection Accuracy | Permissioned Blockchain, IPFS | High energy consumption in WSN environments |
| Hierarchical Incentive Model [13] | IoT data categorized by sensitivity | Efficiency, Security | Permissioned Blockchain, IPFS | Complex incentive model design and upkeep |
| Trusted Data Sharing Framework [15] | – | Security, Access Control | Public Blockchain, IPFS | Latency and scalability challenges |
| BSSN (Adjustable Storage) [16] | Resource-constrained IoT data | Storage Flexibility, System Adaptability | Dynamic Blockchain, IPFS | Complexity in managing adjustable storage |
| Blockchain for IoT Streaming [20] | IoT streaming data | Throughput, Security | Public Blockchain, IPFS | Computational overhead due to synchronization |
| Hybrid Storage Architecture [26] | Synthetic IoT data | Scalability, Latency, Data Redundancy | Hybrid Model (Centralized & Blockchain), IPFS | Coordination complexity between centralized and decentralized parts |
| Modular Consortium Architecture [29] | – | Privacy, Data Integrity | Consortium Blockchain, IPFS | Interoperability challenges, high implementation costs |

**Table 2: Comparison of Blockchain and IPFS-Based IoT Data Storage Techniques (continued).**

| Name of Technique | Dataset Used | Parameters Considered for Evaluation | Type of Blockchain & IPFS Model Used | Drawbacks in the Approach |
|---|---|---|---|---|
| IoT-Enabled Data Sharing [22] | IoT-enabled data sharing | Security, Data Integrity | IOTA Blockchain, IPFS | High energy consumption for data validation |
| IPFS-based Access Control [23] | Secure file transfer data | Access Control, Authentication | Public Blockchain, IPFS | High computational cost for access verification |
| Distributed Storage for Speech Data [24] | Encrypted speech data | Storage Efficiency, Security | Public Blockchain, IPFS | Synchronization delays with large encrypted datasets |
| IPFS Viewer for IoT [30] | IoT surveillance camera data | Data Sharing, Retrieval Speed | Public Blockchain, IPFS | Increased latency in real-time video access |
| Delegation Model for Medical IoT [31] | Medical robotics data | Security, Delegation Efficiency | Private Blockchain, IPFS | Complexity in handling multi-party delegation |
| On-Chain Off-Chain Model [32] | IoT sensor data | Data Integrity, Storage Efficiency | Hybrid (On-chain & IPFS) | Increased complexity in synchronizing on-chain and off-chain data |
| Health Record Sharing [33] | Personal health records | Encryption Efficiency, Data Sharing | Public Blockchain, IPFS | Overhead due to searchable encryption processes |
| IoT Massive Data Management [34] | IoT big data | Scalability, Data Management | Public Blockchain, IPFS | High latency with increasing data volume |
| Industrial IoT Data Security [35] | Industrial IoT data | Data Security, Access Control | Solana Blockchain, IPFS | Complexity in handling industrial-scale datasets |
| Secure Data Sharing [36] | IoT environment | Privacy, Data Sharing | Public Blockchain, IPFS | Latency in handling large-scale data |
| Hybrid IoT Monitoring System [37] | Skin monitoring data | Security, Data Integrity | Hybrid Blockchain, IPFS | Overhead in maintaining hybrid blockchain |
| Fusion Chain [38] | IoT security data | Security, Privacy | Lightweight Blockchain, IPFS | Limited scalability and flexibility |
| Secure Firmware Management [39] | IoT firmware data | Integrity, Security | Private Blockchain, IPFS | Complexity in updating secure firmware |
| IBAM [40] | IoT data for MQTT protocol | Authentication, Privacy | Public Blockchain, IPFS | Authentication overhead in MQTT environments |

**Table 3: Comparison of 2-way Storage and retrieval approaches in blockchain and IoT**

| Medium of Communication | Values Obtained for the Parameters | Security Analysis |
|---|---|---|
| Blockchain and IPFS [2] | Reliability: 99.5%, Data Retrieval Speed: 1.2s | Data Redundancy Checks |
| Blockchain and Lightweight Consensus [4] | Throughput: 150 TPS, Storage Efficiency: 85% | Homomorphic Encryption |
| Blockchain and Smart Contract [5] | Latency: 180ms, Storage Cost: 0.08 ETH/MB | Decentralized Identity Management |
| Blockchain and IPFS [6] | Latency: 200ms, Transaction Speed: 80 TPS | Public Key Cryptography |
| Blockchain and IPFS [7] | Processing Speed: 5GB/s | Blockchain Consensus Mechanisms |
| Smart Contract and IPFS [8] | Data Integrity: 98%, Gas Cost: 0.0015 ETH | Role-Based Access Control |
| Blockchain and IPFS [9] | Gas Cost: 0.002 ETH, Transaction Cost: 0.05 ETH | Encryption, Digital Signature |
| Blockchain and IoT [10] | Scalability: 1000 nodes, Energy Consumption: 50mJ | Zero-Knowledge Proofs |
| Blockchain and IPFS [12] | Detection Efficiency: 92% | Anomaly Detection Algorithms |
| LoRaWAN, Blockchain and IPFS [14] | Transaction Cost: 0.03 ETH, Latency: 220ms | End-to-End Encryption |
| Blockchain and Smart Contract [17] | Latency: 150ms, Throughput: 120 TPS | Hashing, Access Control |
| Smart Contract and IPFS [18] | Storage Cost: 0.1 ETH/MB, Access Time: 1.5s | Merkle Tree Verification |
| Blockchain and IPFS [19] | Security Level: High | Secure Multi-Party Computation |
| Blockchain and IPFS [20] | Streaming Efficiency: 95% | Data Encryption |
| Smart Contract and IPFS [25] | Authentication Speed: 250ms | Authentication Protocols |

Based on the literature survey, Tables 1 and 2 highlight the different techniques used for storage optimization in the IoT using blockchain and IPFS, focusing on the type of blockchain used, the dataset used, the drawbacks in the approach, the parameters used to evaluate the performance of the framework, and the potential scope of improvement in the work. Tables 3,4,5 discuss the existing 2-way communication, storage, and retrieval approaches - the methodology of interfacing between the device and IPFS, the advantages and disadvantages of the approach, the storage and retrieval mechanisms used in the bi-directional approach, and the parameters evaluated along with the values obtained. It can be inferred from the survey that:

- Most approaches face high computational overhead due to the complex encryption and consensus mechanism, which is not feasible for a resource-constrained device such as IoT[20, 29].
- Limited research on enhancing the scalability of large-scale IoT networks works while maintaining security and low latency[12, 24].

- Many proposed methods are platform-specific, thus limiting cross-platform compatibility. There is a need for a standardized blockchain-IPFS protocol tailored for IoT applications and the development of interoperable frameworks[22, 42].
- Many works do not adequately address the efficient retrieval of IPFS data in resource-constrained environments to accelerate the retrieval of IoT devices [19, 24].
- Mitigating advanced threats such as data tampering, rollback attacks, and Sybil attacks in the IoT context is an open challenge, as current methods focus on basic security measures but lack defense against emerging IoT-specific threats[39, 43].
- The lack of a real-time, automated mechanism for dynamic detection and uploading files from the IoT device without manual intervention[5, 18, 27].
- The latency incurred in the data retrieval and storage process is a disadvantage for resource-constrained environments[21, 41].

Table 4: Comparison of Storage and retrieval mechanisms in blockchain-based systems

| Advantages | Storage and Retrieval Mechanism |
| --- | --- |
| Enhances security and ensures efficient retrieval using cryptographic hashing [6] | Hybrid approach combining IPFS and distributed ledger |
| Emphasizes decentralized storage without specific retrieval optimization [8] | Smart contract-enabled decentralized storage |
| Focuses on keyword-based data retrieval without bidirectional interaction [9] | Centralized cloud storage with blockchain indexing |
| Enables real-time file generation and monitoring, eliminating manual intervention [17] | IPFS with blockchain-based CID verification |
| Reduces latency by integrating smart contracts for retrieval authentication [18] | IPFS integrated with Ethereum smart contracts |

Table 5: Comparison of Blockchain and IPFS-Based 2-Way Communication and Storage Approaches

| Methodology | Advantages | Disadvantages | Metrics Obtained |
| --- | --- | --- | --- |
| Middleman IPFS manages transaction processing. Smart contracts handle key management and access control. Hybrid encryption scheme used [4]. | Enhanced security with blockchain and IPFS. Improved scalability by offloading storage to IPFS. Operational efficiency via Middleman IPFS. | Centralization risk in Middleman IPFS. Complexity in hybrid encryption. | Data retrieval time: 2.5s; Throughput: 150 TPS; Collaboration efficiency: 90% user satisfaction |
| Secure healthcare framework integrating IoT, blockchain, and IPFS. Patient data is encrypted and stored in IPFS, with CIDs recorded on a private blockchain [21]. | Ensures tamper-proof storage for healthcare records. Enables secure, decentralized data sharing. Reduces dependency on centralized healthcare servers. | Latency due to blockchain validation. High computational overhead in encryption. | Data access latency: 1.7s; Blockchain transaction cost: 0.0008 ETH; Encryption time: 2.3s |
| IPFS-based authentication model using smart contracts for secure data transfer. CIDs stored on the blockchain with access metadata [23]. | Secure access control with blockchain enforcement. Efficient authentication and data transfer. | Managing smart contract permissions is complex. Potential congestion from blockchain transactions. | Authentication latency: 1.1s; Data retrieval accuracy: 98.5% |
| Ethereum-based DApp using Ganache and Truffle. Node.js server interacts with IPFS for file storage and Web3.js for blockchain transactions [41]. | Ensures data integrity and transparency. Provides user-controlled data sharing. | High resource consumption for blockchain and IPFS nodes. Latency in data retrieval. | Data upload time: 1.2s (1MB); Access control time: 0.8s; User satisfaction: 85% |
| Digital evidence preservation using IPFS and blockchain. Smart contracts ensure access control and integrity verification of forensic data [42]. | Reliable and tamper-proof evidence storage. Secure access control. Traceability of modifications. | High storage costs for forensic data in blockchain. Scalability issues for large-scale deployments. | Retrieval time: 1.9s; Smart contract time: 0.75s; Storage cost/MB: 0.0005 ETH |

## 3 Experimental Setup and Methodology

This section outlines the experimental setup and methodology for the optimized data storage and retrieval framework, leveraging IPFS, Python, and blockchain technologies. The first phase is a scalable 2-way framework that enables real-time data storage and retrieval between IoT and IPFS using Ethereum smart contracts for automated file monitoring, CID management, and dynamic retrieval using the IPFS gateway. The second phase optimizes further by developing a blockchain-inspired lightweight framework that integrates IPFS with a custom Python blockchain to store the CIDs and use them for data fetching when requested by the user, and the use of SHA-256 hashes for preventing data tampering and ensuring data verification and integrity. Sections 3.1 and 3.2 describe the

3.1 Phase 1: A real-time, scalable, and optimized data storage and retrieval framework using IPFS and Smart Contracts.

We have considered the Raspberry Pi 4 model B as our resource-constrained device and have developed a custom environment, mirroring the Raspberry Pi 4 model B device with a 64-bit processor and 8GB of RAM. The various software and libraries used for the proposed approach are listed below:

1. IPFS version 0.29.0
2. Node JS version 20.16.0
3. Geth (Go Ethereum) version 1.13.15
4. Remix IDE for deploying smart contracts

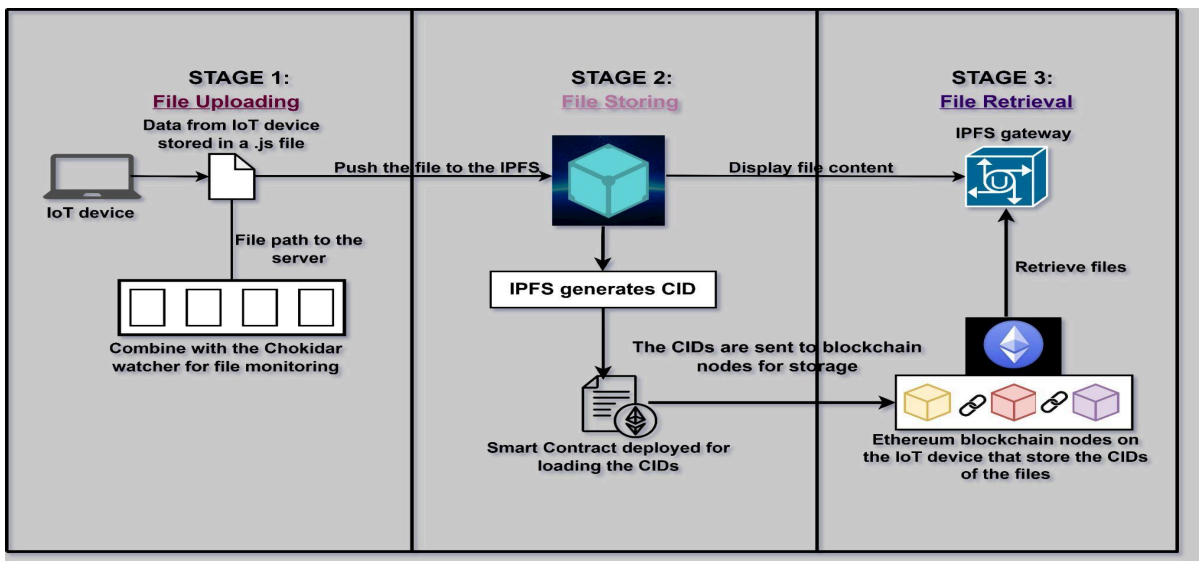

Fig. 2: The architecture diagram of the 1$^{st}$ phase on a scalable and real-time data storage and retrieval framework using IPFS and smart contracts.

Figure 2 shows the architecture diagram for our proposed approach, which involves 3 stages and is explained through Algorithm 2-the 1$^{st}$ stage involves the data monitoring for new files on the device and simultaneous uploading of the detected files to the IPFS. After the file is successfully uploaded to the IPFS, the corresponding CIDs are generated, and smart contracts are deployed to connect with the IPFS client and fetch the CIDs in the 2$^{nd}$ stage. Finally, in the 3$^{rd}$ stage, the CIDs are sent from the IPFS to the Ethereum nodes via smart contracts, and the IPFS gateway is used for the retrieval of files when requested by the user via the IPFS client and smart contracts. The data set used for the experiments was dummy files generated from JavaScript in real-time with a .dat extension, and 15 in number that were saved on the system - the minimum and maximum file sizes were set to 5MB and 550MB, respectively. The reason for using .dat files lies in the fact that these format files can handle any unstructured or structured data, as in IoT devices, and are faster to generate and incur low computational overhead that meets the requirements for a resource-constrained device.

The pseudo-code for the same is given as in Algorithm 1; Algorithm 3 describes the procedure of uploading the files to IPFS and the simultaneous check on the directory for any new files saved that need to be uploaded to IPFS. Similarly, Algorithm 4 describes the smart contract code written for the metrics recorded for storing the CIDs obtained from the IPFS to the blockchain node for estimating the latency of our approach. The metrics recorded, mainly the energy consumption, system resource utilization, and the data storage procedure on IPFS, have been described in Algorithm 5. The results have been described in the Results and Analysis section.

**Algorithm 1: Generate Dummy Files with Increasing Gaps**

Input: Initial file size $S$ = 5 MB, Gap $G$ = 5 MB, Maximum file size $M$ = 550 MB
Output: Series of files with increasing sizes
Create directory generated files if it does not exist;
currentSize ← $S$;
while *currentSize* ≤ $M$ do
    fileName ← "dummy file " + currentSize + "MB.dat";
    filePath ← generated files + fileName;
    Create a file at filePath filled with zeros of size currentSize MB; Print "Created fileName of size currentSize MB";
    currentSize ← currentSize + gap;
    gap ← gap + 5;

**Algorithm 2: File Storage and Retrieval using IPFS and Blockchain for a 2-way Communication Approach**

Input: IoT device data, IPFS storage, Ethereum blockchain
Output: Stored file CID, blockchain registration, retrieval capability Stage
1: File Uploading;
Initialize IPFS and start the daemon on localhost:5001;
Set up the Ethereum environment using Ganache or Geth;
Initialize a Node.js project for file management;
Monitor the directory for new files using Chokidar (every 20 minutes); if *a new file is detected* then
    Read the file from the local server;
    Push the file to the IPFS network;

Stage 2: File Storing;
Generate a CID for the file after successful upload to IPFS;
Deploy a Solidity smart contract on Ethereum for CID storage;
Store the CID in the Ethereum blockchain by interacting with the deployed smart contract;
Confirm storage by verifying the CID on the Ethereum node;
Stage 3: File Retrieval;
Retrieve the CID from the Ethereum blockchain using the smart contract; Fetch the file from the IPFS gateway using the CID;
if *retrieval is successful* then
    Display the retrieved file content;
    Validate integrity by comparing with the original file hash (if needed); End
Process

**Algorithm 3: Automated File Upload to IPFS using Chokidar Require:**

Chokidar, IPFS node, and file system access
1: Initialize IPFS client at http://localhost:5001/api/v0
2: Define function formatFileSize(bytes):
3: Set units = [Bytes, KB, MB, GB]
4: Initialize $i$ ← 0
5: Repeat:
6: bytes ← bytes / 1024
7: $i$ ← $i$ + 1
8: Until: bytes < 1024 or $i$ = units.length - 1
9: Return bytes.toFixed(2) + " " + units[i]
10: Initialize Chokidar watcher on folder:
    C:/Sem 5/Minor project/Chokidar test/CCTV
11: Set watcher to ignore hidden files and keep persistent monitoring 12:
On event: File Added

13: Read file content into fileContent
14: Get file size using fs.statSync()
15: Record start time as startTime
16: Upload file to IPFS using ipfs.add(fileContent)
17: Record end time as endTime
18: Compute upload time = endTime - startTime
19: Print CID, file size, and upload time
20: Handle and log errors during upload
21: Print message: "Watching for file changes..."
22: Set timeout to stop watching after 20 minutes
23: On timeout: close watcher and print
"Stopped watching after 20 minutes."

Algorithm 4: IPFS Storage Smart Contract Algorithm
Require: Ethereum blockchain, IPFS node
1: Define struct Data with attributes:
2: ipfsHash (string) – IPFS hash of stored data
3: timestamp (uint256) – Time when data was stored
4: storeTimeTaken (uint256) – Time taken to store the data
5: Define mapping storedData to store user data lists
6: Define events:
7: DataStored(user, ipfsHash, timestamp, storeTimeTaken) 8: DataRetrieved(ipfsHash, timestamp, retrievalTimeTaken) 9: Function: storeData( ipfsHash: string)
10: startTime ← block.timestamp
11: timestamp ← block.timestamp ▷ Simulate store time 12: storeTimeTaken ← timestamp - startTime
13: Append Data( ipfsHash, timestamp, storeTimeTaken) to storedData[msg.sender]
14: Emit DataStored(msg.sender, ipfsHash, timestamp, storeTimeTaken) 15: Function: retrieveData(index: uint256)
16: Require: index < length(storedData[msg.sender])
17: data ← storedData[msg.sender][index]
18: retrievalTimeTaken ← block.timestamp - data.timestamp 19: return (data.ipfsHash, retrievalTimeTaken)
20: Function: getDataCount()
21: return length(storedData[msg.sender])

Algorithm 5: Upload Files to IPFS and Log Energy Consumption Input: Folder path $F$, Output CSV file path $C$, CPU TDP in Watts $TDPWATTS$
Output: Each file in $F$ is uploaded to IPFS, and energy consumption is logged in $C$
Function CalculatePower(*cpu usage*):
    return (*cpu usage*/100) × $TDPWATTS$;
Function UploadFilesAndLogPower(*folder path, output csv path*):
    Connect to local IPFS node;
    Get list of files in $F$;
    if *no files found* then
        Print "No files found in the specified folder";
        return;
    Open CSV file $C$ and write header;
    foreach *file in F* do
        *file size* ← size of file in MB;
        *start time* ← current time;
        *initial cpu* ← CPU usage before upload;
        Print "[INFO] Upload started for file at start time";
        Upload file to IPFS;

    *cid* ← CID of uploaded file;
    *end time* ← current time;
    *final cpu* ← CPU usage after upload;
    *duration* ← *end time* − *start time*;
    *avg cpu* ← (*initial cpu* + *final cpu*)/2;
    *power consumed* ← CalculatePower(*avg cpu*) ×*duration*;
    Print file details, CPU usage, and energy consumption;
    Write results to CSV file;
  Print "Power report saved to output CSV file";
 Call UploadFilesAndLogPower(*Folder Path, Output CSV Path*);

Table 6 shows the various metrics that have been used for evaluating and recording the performance of the proposed approach, the main metrics being the Time taken to upload the file to the IPFS, the memory consumption, the power consumption, the retrieval time for the file, and the amount of Gas required to deploy the smart contracts.

**Table 6: Definitions and Equations for the Parameters Used in Experiment**

| Metric | Explanation | Equation | Unit |
|---|---|---|---|
| Time to Upload | Time taken for full upload (processing + transmission). | $T_{upload} = T_{processing} + T_{network}$ | Seconds (s) |
| CPU Usage | Processor use during operation. | $CPU_\% = \frac{CPU_{used}}{CPU_{total}} \times 100$ | % |
| Memory Usage | RAM used during the operation. | $M_{used} = M_{final} - M_{initial}$ | MB |
| Power Consumption | Electrical energy consumed. | $P = V \times I \times T$ | Joules (J) |
| Gas Used (Deploy) | Gas used to deploy the contract. | $G_{deploy} = \sum G_{op}$ | Gas units |
| Gas Cost | Blockchain cost in ETH or Gwei. | $G_{cost} = G_{used} \times G_{price}$ | ETH / Gwei |
| Transaction Cost | Complete blockchain TX cost. | $C_{tx} = G_{used} \times G_{price}$ | ETH / Gwei |
| Retrieval Time | Time taken to download + reconstruct. | $T_{retrieve} = T_{lookup} + T_{network}$ | Seconds (s) |
| Transaction Hash | Unique ID for blockchain TX. | Not applicable (hash output) | Alphanumeric |

### 3.2 Phase 2: A blockchain-inspired lightweight Python-based framework for data storage and retrieval to and from IPFS.

To further optimize the framework in Phase 1 with respect to file upload and retrieval time, we designed a Python-based blockchain framework that poses as an alternative to the traditional blockchain, where the computational burden to deploy a blockchain increases the computational overhead for the IoT device in use.

For setting up the blockchain and IPFS environments, the experiments were performed with the same custom environment as used in Phase 1, mimicking the specifications of a Raspberry Pi 4 model B device: 64-bit processor and 8GB RAM. Image and text files were used for the experimental setups that were stored on the device, and the file size ranges from 87 KB to 5 MB. The following software was used for the experimental setup with the corresponding version numbers:

- Python - Jupyter Notebook - version 3.12.7
- IPFS version 0.34.1 running on https://127.0.0.1:5001/webui(localhost), and the files to be retrieved are available on port 8080 (the gateway).

Figure 3 shows the flow diagram of our proposed approach, which has been described below:

- The device selects a file to be uploaded to the IPFS.
- The file is provided as input to the Python program for upload, and is transmitted using the IPFS version 0.34.1.
- The file is pinned to ensure it remains available on the IPFS host.
- After a successful upload, IPFS generates a CID that is the unique identifier for the uploaded data.
- A new block structure is created using Python to store the CIDs obtained for the uploaded files.
- The file corresponding to the CID is fetched from IPFS to validate its authenticity and tamper-proofing.

Algorithms 6 and 7 present the pseudo-codes used to perform the experiments, divided into 2 parts as in Figure 3.

Algorithm 6: IPFS File Upload and Retrieval
1: procedure UploadFile(file path)
2:  ipfs path ← "D:/kubo/ipfs.exe"
3:  Print "Uploading..."
4:  result ← Run ipfs add file path
5:  if result.returncode $\neq$ 0 then
6:   Print "Upload failed"
7:   return None
8:  cid ← result.stdout (trimmed)
9:  Print "CID: ", cid
10: return cid
11: end procedure
12: procedure RetrieveFile(cid, output path)
13: ipfs path ← "D:/kubo/ipfs.exe"
14: Print "Retrieving..."
15: result ← Run ipfs cat cid > output path
16: if result.returncode $\neq$ 0 then
17:  Print "Retrieval failed"
18:  return False
19: Print "Saved to: ", output path
20: return True
21: end procedure
22: procedure Main
23: input ← "C:/Users/sanja/retrieved file.png"
24: output ← "downloaded retrieved file.png"
25: cid ← UploadFile(input)
26: if cid $\neq$ None then
27: RetrieveFile(cid, output)
28: end procedure
29: Main

Algorithm 7: Fetching Data from IPFS and Storing in Python Blockchain 1:
   Procedure FetchData(cid)
  2: result ← ipfs cat cid
  3: if result.success then
  4: return result.stdout
  5: else
  6: return None

  7: Procedure SaveToChain(cid)
  8: data ← FetchData(cid)
  9: if data ≠ None then
 10: hash ← SHA-256(data)
 11: prev ← blockchain.last_block().hash
 12: blockchain.add_block(prev, cid, hash)

 13: Procedure VerifyChain()
 14: for each block in blockchain do
 15: data ← FetchData(block.cid)
 16: if data ≠ None and SHA-256(data) = block.hash then 17: Print "Verified"
 18: else
 19: Print "Tampered"

 20: Procedure Main()
 21: blockchain ← Init new blockchain
 22: for each cid in {cid 1, cid 2, ...} do
 23: SaveToChain(cid)
 24: VerifyChain()

**Fig. 3 The proposed workflow of the second phase, the Lightweight Python-based blockchain framework for IPFS data storage and retrieval.**

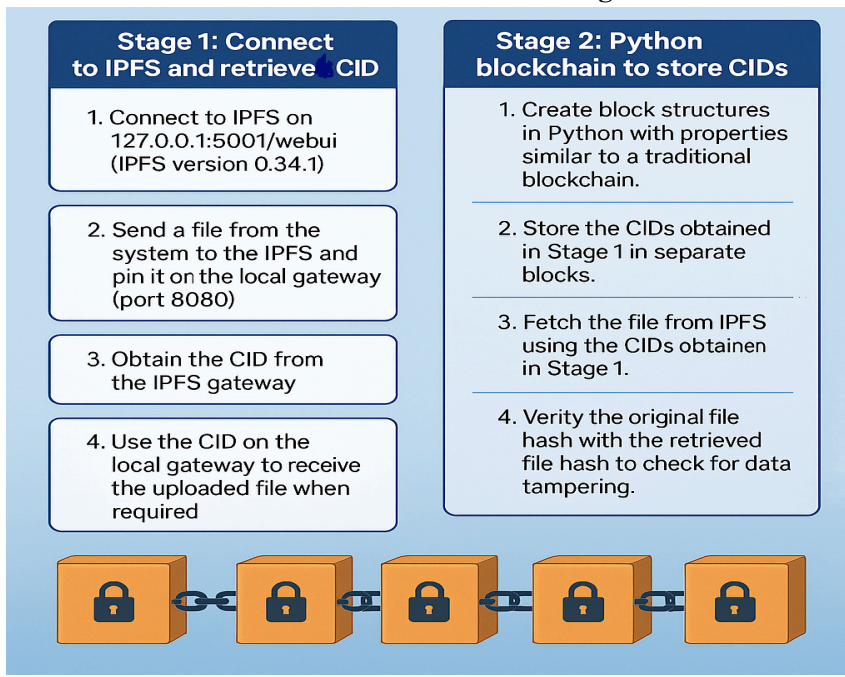

# 4 Results and Analysis

This section discusses the metrics and the values obtained for the experiments performed for the 2 phases with detailed analysis, graphical visualizations, and the performance parameters recorded. Sections 4.1 and 4.2 describe the results obtained for the 2 phases, respectively. Table 6 illustrates the various performance metrics obtained for the experiments, as shown through Algorithms 1,2,3,4,5,6, and 7.

## 4.1 The metrics recorded for the $1^{st}$ phase on a scalable and real-time data storage and retrieval framework using IPFS and smart contracts

Some of the general analysis, pre-assumptions, and the constraints/challenges faced while performing the experiments were:

• The time taken to upload a file increases linearly with the increase in file size. • Node.js has a buffer limit of 1.0 GB in a 32-bit architecture and 2.0 GB in a 64-bit architecture, so the maximum file size that could be uploaded was 2.0 GB. To tackle this issue, the files could be broken down into smaller chunks so that the effective size is below 2.0 GB or 1.0 GB, depending on the host architecture. • Persistent watching is enabled in Chokidar by default, which uses more resources than necessary.

• The files could not be copied and pasted without throwing an error, and could only be moved. This is because when Chokidar watches over the directory while the file is being copied, the file is modified and used at that instance.

• To tackle the above-mentioned issues, we configured Chokidar to scan the directory at periodic intervals instead of continuously. This saved resources and greatly reduced the chances of the aforementioned errors occurring.
• A set data structure to store the details of already uploaded files was used to avoid re-uploading files.
• Since Chokidar is highly configurable, further changes as per requirements are possible.

### 4.1.1 The metrics recorded to upload the files from the device to the IPFS

Table 7 and Figure 4 display the values recorded when the files are uploaded to the IPFS: the time to upload the file, the energy, and the memory consumption, respectively. Figure 5 is a 3D visualization of the variations obtained for different file sizes against different performance metrics.

**Table 7: System metrics recorded during file uploads to IPFS, including uploading time, power consumption, and memory usage.**

| File Size (MB) | Uploading Time (s) | Power Consumption (J) | Memory (MB) |
|---|---|---|---|
| 5.0 | 0.23 | 2.57 | 66.62 |
| 10.0 | 0.49 | 4.59 | 86.90 |
| 20.0 | 0.68 | 17.41 | 137.11 |
| 35.0 | 0.94 | 9.35 | 197.18 |
| 55.0 | 0.96 | 12.31 | 266.83 |
| 80.0 | 1.45 | 19.67 | 366.92 |
| 110.0 | 1.77 | 35.28 | 486.79 |
| 145.0 | 2.38 | 37.86 | 626.97 |
| 185.0 | 2.97 | 43.62 | 787.08 |
| 230.0 | 3.19 | 53.35 | 967.20 |
| 280.0 | 3.96 | 62.03 | 1167.29 |
| 335.0 | 4.58 | 76.53 | 1365.38 |
| 395.0 | 5.44 | 88.20 | 1605.93 |
| 460.0 | 6.28 | 121.19 | 1866.20 |
| 530.0 | 7.63 | 136.94 | 2146.53 |

For the energy consumption obtained(Figure 4), it can be inferred that:

**Fig. 4 The graphic visualization for the time taken, power consumed, and the memory utilized to upload a file to IPFS.**

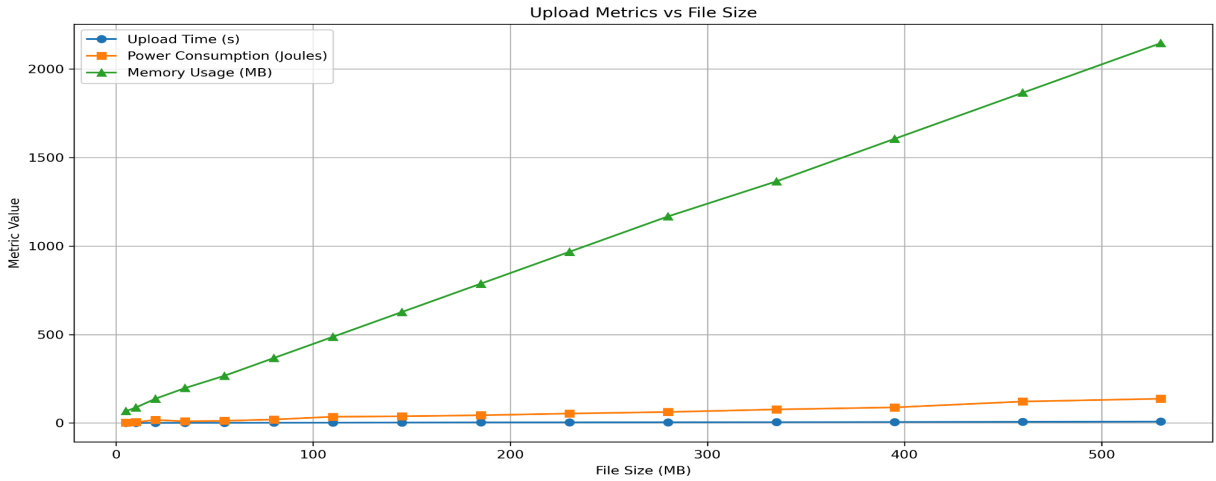

- Smaller files experience fluctuating power usage due to higher relative overheads from node initialization, hashing processes, and disk I/O.
- CPU usage spikes noticeably while uploading small files, directly impacting energy consumption.
- For files larger than 100 MB, energy consumption scales more predictably with file size, as overheads become less significant compared to the time required to process larger payloads.
- Energy consumption becomes more consistent as file size increases, with a clear correlation between CPU usage and file size.
- During the upload of smaller files, the IPFS node undergoes initialization, including hashing algorithms and network setup. This results in spikes in CPU usage and, subsequently, power consumption.
- Delays in accessing and writing files, especially when using slower storage (e.g., HDD), lead to inefficient CPU usage, further affecting energy consumption. • Background tasks such as garbage collection, resource cleanup, or other system-level processes can interrupt the file upload process, causing temporary CPU usage and power consumption spikes.

The analysis of the time taken to upload the file(Figure 4) states that:

- Files smaller than 50 MB experience fluctuating upload times due to the proportionally larger impact of overheads.
- Overheads include initialization of the IPFS node, hashing processes, and HTTP connection setup.
- Beyond 100 MB, the upload time increases predictably with file size, as fixed overheads become negligible compared to the processing time for larger payloads. • The initial setup of the IPFS node, hashing algorithms, and HTTP connection delays for uploads of smaller files.
- Delays in accessing and writing files to disk, including caching effects, impact the performance for smaller chunks.

- Background processes, such as garbage collection or resource cleanup, can temporar ily slow the upload process.

For the memory consumed during the file upload(Figure 4), the following inferences are made:

- The graph shows a linear relationship between file size and memory usage, with memory consumption increasing proportionally as the file size grows. • Smaller files (e.g., 10 MB) use minimal memory (86.9 MB), while larger files (e.g., 530 MB) require significantly more memory (2146.5 MB).
- The trend suggests predictable system behavior, making resource planning easier. • For memory-constrained systems (e.g., 32-bit architecture), large files may exceed memory limits, requiring optimization techniques like chunked processing. • The graph highlights the need for efficient resource management for handling large files without compromising system stability.

### 4.1.2 The metrics recorded for storage on the blockchain node and retrieval from the IPFS gateway

Tables 8 and 9 show the metrics obtained for the data retrieval between the blockchain node(Ethereum node) and the IPFS gateway, and the status of the data stored on the Ethereum nodes for the data sent from the IPFS via a smart contract. The meaning of the metrics recorded with their analysis is given below:

1. Block Number represents the sequential identifier of a block in the blockchain, containing a group of transactions validated and recorded at a specific time.

2. Transaction Hash is a unique identifier (hash) for a transaction used to verify and track its status and details within the blockchain.

3. IPFS Hash/CID is a unique content identifier generated by IPFS that links data stored off-chain and ensures immutability in smart contracts.

4. User Address is a unique alphanumeric string representing a user's identity in the blockchain, often used to initiate or receive transactions.

5. Timestamp is the exact time a block was mined or a transaction occurred, helping ensure chronological order and transparency.

6. Store Time is the time taken to upload and record data (e.g., IPFS hash) on the blockchain, indicating storage efficiency as the time taken is almost negligible. 7. Gas Cost is the amount of cryptocurrency paid to miners for executing and recording a transaction or smart contract on the blockchain, reflecting computational effort, the contract complexity, and network congestion.

Tables 8 and 9 show the gas and transaction costs incurred, that is less in comparison to the works done in [8, 9, 14, 21, 42], indicating that our approach of communication between the device and IPFS via smart contract is optimized concerning the gas fees, the latency, and the memory and energy consumption, which is suitable for a resource-constrained environment like the IoT.

**Fig. 5** The 3D visualization for the metrics calculated for uploading the different file sizes to IPFS.

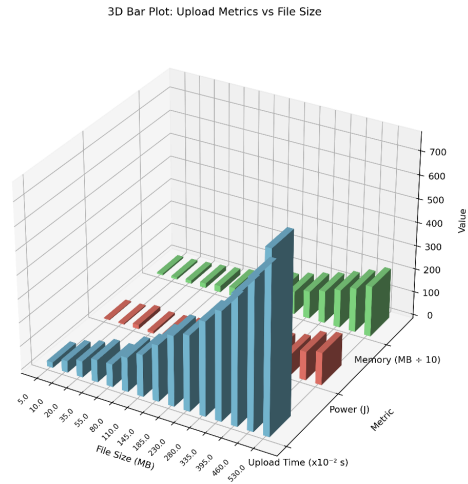

It is visible through Algorithms 3 and 4 that the procedure opted for is simple with minimum operations, specifically for the smart contract code that resulted in low gas and transaction costs.

- The CIDs are stored on the nodes, not the complete data, which resulted in low gas cost.
- There is minimal interaction with the blockchain as the data is retrieved upon request, thus further reducing the cost.
- The smart contract codes are written that are simple and require minimal computations for interaction with the IPFS and the blockchain nodes, thus incurring less gas and transaction costs.

**Table 8: The metrics for Blockchain to IPFS Mapping.**

| Block Number | Transaction Hash | Gas Cost Units | Transaction Cost Units | Retrieval Time (ms) | IPFS Hash |
|---|---|---|---|---|---|
| 3157 | 0x2742d83...8e07ac909 | 752110 | 654008 | 85 | Qmbt4P2...XEKvuk |
| 3159 | 0xb604211...0a119ad12 | 752110 | 654008 | 75 | QmRt5rhjt...6nArLocv |
| 3161 | 0x8a9b2c8...34ce4ae3 | 752110 | 654008 | 65 | QmajztYQ...yk8pDT |

**Table 9: Metrics recorded for data storage on the blockchain nodes.**

| Block | Transaction Hash | IPFS Hash | User Address | Timestamp | Store Time (ms) |
|---|---|---|---|---|---|
| 1298 | 0xc3f17d93...db8b357b | Qmdr9cJs...C46ei | 0x24D36Be...421Fff | 61724740207 | 75 |
| 1299 | 0xcd911f7e...3efea | QmNyj9A2...Y26y | 0x24D36Be...421Fff | 61724740222 | 110 |
| 1300 | 0x4253db29...597f490a | QmQdGkKv...wpHQk | 0x24D36Be...421Fff | 61724740237 | 105 |

## 4.2 Comparison with the existing work

Table 10 and Figure 6 compare our work with the existing work, the time taken to upload a file and the gas units consumed (inferred from Tables 8, 9, and 10), and the security aspect of the proposed approach. In contrast to the approach presented in [9, 12, 17, 18, 44], our proposed methodology significantly improves several areas. While their solution effectively combines blockchain, IPFS, and encryption techniques to ensure data immutability, traceability, availability, and privacy, our approach enhances these aspects by implementing a more efficient data retrieval mechanism, reducing latency, and improving user experience. Additionally, we have optimized the encryption process to minimize computational overhead, thereby increasing the system's overall scalability and performance. These advancements position our solution as a more robust and efficient alternative for secure distributed data storage and retrieval.

**Fig. 6 The upload time and file size for various works.**

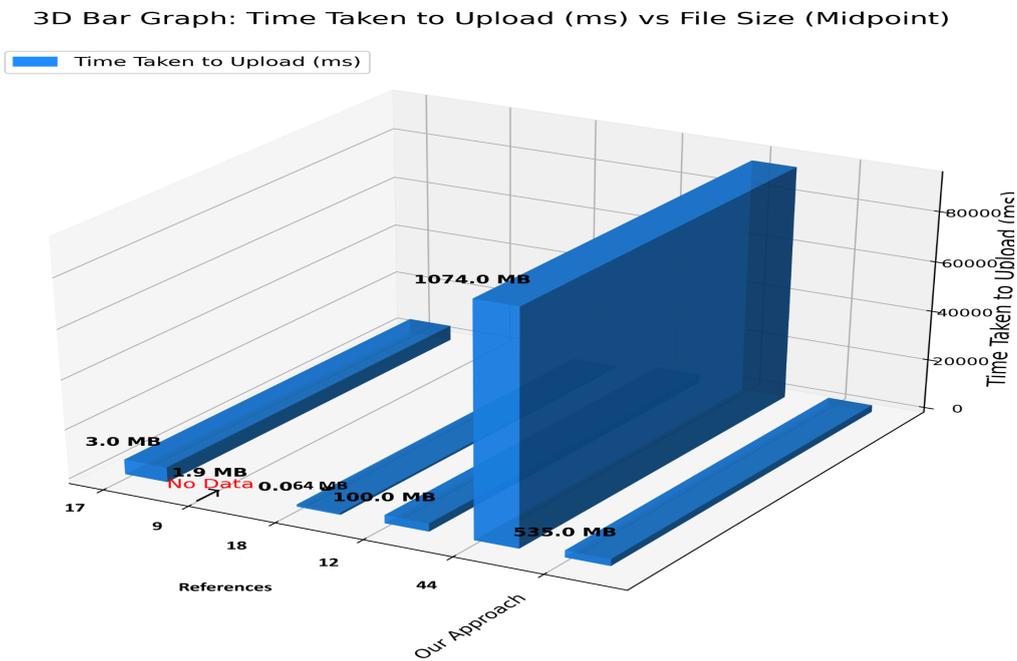

**Table 10: Comparison of Different Blockchain and IPFS-Based Storage Approaches.**

| Metrics/Reference | [17] | [9] | [18] | [12] | [44] | Our Approach |
|---|---|---|---|---|---|---|
| File Size | 10KB–3MB | 898KB–1MB | 3KB–11.7KB | 1KB–100MB | 50MB–1GB | 5MB–530MB |
| Time to Upload (avg) (ms) | 6000 | – | 859.7 | 3300 | 93700 | 2800 |
| Blockchain Used | Ethereum | Ethereum | Ethereum | Ethereum | EOSIO | Ethereum |
| Gas Units Consumed | – | 1047485 | 144126 | 1000000 | – | 752110 |

**Fig. 7: The gas consumption across various works**

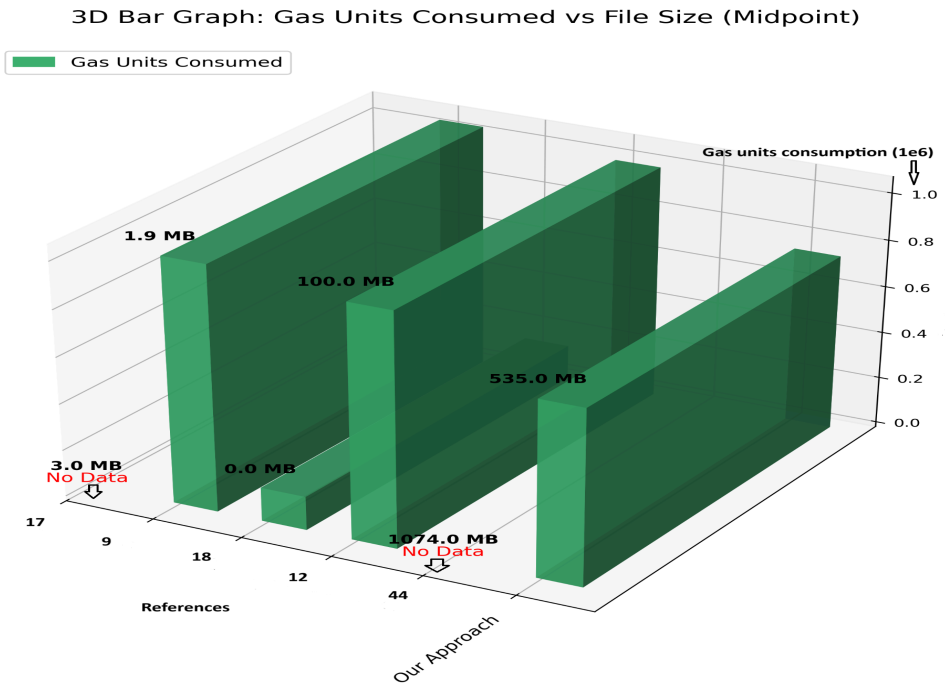

4.3 The metrics recorded for the 2nd phase on a lightweight data storage and retrieval using Python-blockchain and IPFS

Table 12 displays the performance metrics obtained for data upload and retrieval using
IPFS. The significance of the metrics evaluated is defined below:
• Content Identifier (CID): It is a unique, immutable cryptographic hash generated
by IPFS for each file uploaded to the system. The CID changes if any change is detected in the content of the file, thereby making it also content-addressable.

$$CID = Hash(File\ contents) \quad (1)$$

The hash is a secure algorithm like SHA-256, 32 bytes in length.
• Memory Usage refers to the amount of RAM consumed during an operation,
denoted in MB. It signifies the computational load on the system when the program
is in execution.
• Bandwidth refers to the data transfer speed during the upload and retrieval,
expressed in KB/s or MB/s as represented in (2).

$$Bandwidth = File\ Size\ (KB\ or\ MB)/Time\ Taken\ (s) \quad (2)$$

• Time to upload is the total duration to send and upload the file to IPFS.
• Time to retrieve measures the duration required to fetch the file from IPFS using
the CID.
Figure 7 shows the graphic visualizations of the comparison of our approach with the
existing approaches in terms of file size and total time taken to upload and fetch from
the IPFS, which indicates the efficiency of our approach. Based on Tables 11 and 12,
it can be inferred that:
• The proposed approach demonstrated a fast upload and retrieval process, implying
efficient data handling.
• A high throughput was obtained as seen from the bandwidth values, implying that
fast data exchange and communication happened efficiently in less time.
• The memory usage was quite low, indicating the efficient utilization of system
resources and the low complexity of the proposed code.
• The block storage performance shows fast fetch time, and a reliable verification
mechanism through the stored and fetched hashes to prevent data tampering and data integrity.
• The blocks created using Python for the experiment are reliable and linked via cryptographic hashes; also, the identifiers like the timestamp, the previous, and current hashes help in maintaining data traceability and immutability.

Table 13 displays the comparison of our proposed blockchain-inspired Lightweight Python-based framework based on the essential characteristics required in a blockchain and its fulfillment in our proposed blockchain; it can be inferred from the table that Our proposed Python blockchain is able to provide the necessary features required for secure data storage, and through the experimental results, shows its suitability for deployment in resource-constrained environments. From the comparison table in Table 14, it can be inferred that our approach demonstrated efficient and lightweight data upload and retrieval time, in addition to the type of storage structure used (Python in our case) that can be deployed in resource-constrained devices without extra computational load.

Fig. 8: The comparison of the file upload time for various works with our proposed approach.

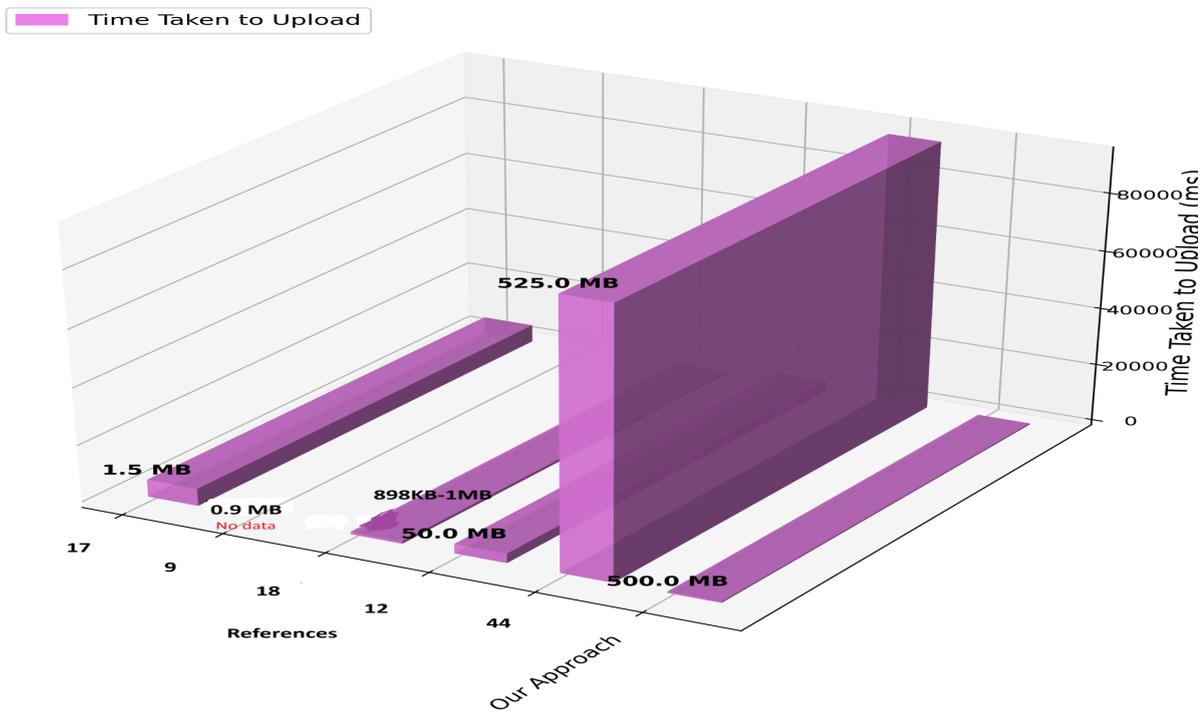

Table 11: Performance metrics for data upload and retrieval using IPFS.

| Operation | Time (s) | Memory (MB) | Bandwidth (KB/s) | Size (MB) | CID (truncated) |
|---|---|---|---|---|---|
| Upload | 0.15 | 0.02 | 7558.58 | 1.1 | QmZQX...E71784RC |
| Retrieval | 0.13 | 0.00 | 8777.44 | 1.1 | QmZQX...E71784RC |
| Upload | 0.24 | 1.21 | 31956.85 | 5.1 | QmSoj...iCm2KTZJ |
| Retrieval | 0.16 | 0.05 | 21616.15 | 5.1 | QmSoj...iCm2KTZJ |
| Upload | 4.20 | 0.16 | 119112.77 | 500 | QmSxx...V1dvucwm |
| Retrieval | 7.37 | 0.01 | 67837.65 | 500 | QmSxx...V1dvucwm |

Table 12: Performance metrics for files stored on the block.

| Block | CID (truncated) | Time (s) | Size (B) | Stored Hash (truncated) | Fetched Hash (truncated) |
|---|---|---|---|---|---|
| 0 | – | – | – | – | – |
| 1 | QmX2mnf7kU...uULNP | 10.48 | 133397 | 7013d64921...8854f6c | 7013d64921...8854f6c |
| 2 | QmTTQ5Xpwd...Phnoz3 | 10.11 | 797553 | 118f552937...512b1e2 | 118f552937...512b1e2 |
| 3 | QmZWDbHdSU...LZVW6 | 10.41 | 152 | 3c0d3b7ec9...5effb | 3c0d3b7ec9...5effb |
| 4 | QmdQGKkvPD...wpHQJ | 10.07 | 87 | fb5cc3fdfe...8425d | fb5cc3fdfe...8425d |

Table 13: Validation of Blockchain Properties in our proposed Python Blockchain.

| S.No. | Test Criteria | Purpose | Validation in Python Blockchain | Available? |
|---|---|---|---|---|
| 1 | Linked list of blocks | Ensure immutability and order | Each block includes hash of previous block | Yes |
| 2 | Block content integrity | Secure data encapsulation | Block has timestamp, data hash, and IPFS CID | Yes |
| 3 | Cryptographic hashing | Tamper-resistance | SHA-256 used for block hash | Yes |
| 4 | Consensus mechanism | Block validation agreement | Not implemented (single node) | Partial |
| 5 | Immutability | Detect tampering | Hash chain breaks on any change | Yes |
| 6 | Timestamping | Maintain order | UTC timestamp stored in each block | Yes |
| 7 | Decentralization | Peer-to-peer replication | Not implemented | No |
| 8 | Append-only structure | Prevent changes | Blocks only appended | Yes |
| 9 | Data validation | Avoid malicious data | Data hash checked before adding | Yes |
| 10 | Chain consistency | Consistency across nodes | Not applicable (single node) | No |

Table 14 Comparison of our approach with existing blockchain and IPFS-based approaches.

| Metric / Reference | [17] | [9] | [18] | [12] | [44] | Our Approach |
|---|---|---|---|---|---|---|
| File Size | 10KB–3MB | 898KB–1MB | 3KB–11.7KB | 1KB–100MB | 50MB–1GB | 500MB |
| Upload Time (avg) (ms) | 6000 | – | 859.7 | 3300 | 93700 | 4200 |
| Retrieval Time (avg) (ms) | – | – | 1263 | 2671 | – | 5780 |
| Blockchain Used | Ethereum | Ethereum | Ethereum | Ethereum | EOSIO | Python Blockchain |

## 5 Conclusion and Future Work

This article presents an optimized framework for secure and decentralized data storage and retrieval in the IoT using IPFS, smart contracts, and Python through progressive phases for various application needs. In the first phase, we developed a scalable 2-way data storage and retrieval framework, deploying real-time file monitoring for data upload to IPFS and combining with Ethereum smart contracts for dynamic retrieval and achieving low file upload time, making it feasible for real-time and critical IoT networks. The second phase further optimizes the results obtained in the first phase through a lightweight Python-based blockchain-inspired framework for data storage and retrieval that integrates IPFS with Python to store CIDs and use them for data retrieval when requested by the user. Unlike Ethereum and similar blockchain platforms that incur significant latency and resource overhead, this Python-based blockchain provides a faster file upload time and low memory usage. It also employs SHA-256 for ensuring data integrity; the entire framework is robust and achieves fast upload and retrieval time, as verified by the experimental results. Future work can enhance the lightweight blockchain framework for dynamic block creation, with data encryption and decryption for increasing data security and privacy, and for protection against Sybil attacks by enhancing the smart contracts framework. Also, a hybrid framework can be built by combining these 2 frameworks and incorporating access control mechanisms to validate its utility in decentralized storage mechanisms, real-time IoT applications, and resource-constrained critical devices.